\documentstyle[aps,prd,12 pt]{revtex}

\tightenlines

\newcommand{\beq}{\begin{equation}}
\newcommand{\eeq}{\end{equation}}

\input epsf

\begin{document}

\title{Quantum cosmology and eternal inflation}

\author{Alexander Vilenkin}

\address{Institute of Cosmology, Department of Physics and Astronomy,\\
Tufts University, Medford, MA 02155, USA}

\maketitle

\begin{abstract}

This contribution consists of two parts.  In the first part, I review
the tunneling approach to quantum cosmology and comment on the
alternative approaches.  In the second part, I discuss the relation
between quantum cosmology and eternal inflation.  In particular, I
discuss whether or not we need quantum cosmology in the light of
eternal inflation, and whether or not quantum cosmology makes any
testable predictions.

\end{abstract}

\section{Introduction}

Stephen and I at times disagreed about minor things, like the sign in
the wave function of the universe, $\psi \sim e^{\pm S}$.  But for me
Stephen has always been a major source of inspiration.  I will mention
just one episode, when I talked to Stephen at a conference and was
telling him why I thought my wave function was better than his.  To
which Stephen replied: ``You do not have a wave function.''

Talking to Stephen is a little like talking to the Oracle of Delphi:
you are likely to hear something as profound as it is difficult to
decipher.  But in that particular case I knew immediately what he
meant.  He was pointing to the lack of a general definition for the
tunneling wave function, which at the time was defined only in a
simple model.  My talk was scheduled for the next morning, so I spent
the night working out a general definition and rewriting my
transparencies.  I ended up giving a very different talk from what was
initially intended.  I can thus say with some justification that
Stephen contributed to the development of the tunneling approach,
although he may not be very pleased with the result.

My contribution here will consist of two parts.  In the first part, I
will review the tunneling approach to quantum cosmology and will
briefly comment on the alternative approaches.  In the second
part, I will discuss the relation between quantum cosmology and
eternal inflation.  After a brief review of eternal inflation, I will
discuss whether or not we need quantum cosmology in the light of
eternal inflation, and then whether or not quantum cosmology makes any
testable predictions.

\section{Quantum cosmology}

If the cosmological evolution is followed back in time, we are driven to
the initial singularity where the classical equations of general
relativity break down.  There was initially some hope that the
singularity was a pathological feature of the highly symmetric
Friedmann solutions, but this hope evaporated when Stephen and Roger
Penrose proved their famous singularity theorems.  There was no
escape, and cosmologists had to face the problem of the origin of the
universe. 

Many people suspected that in order
to understand what actually happened in the beginning, we
should treat the universe quantum-mechanically and describe it by a
wave function rather than by a classical spacetime.  This quantum
approach to cosmology was initiated by DeWitt \cite{DeWitt} and Misner
\cite{Misner}, and after a somewhat slow start received wide recognition
in the last two decades or so.  The picture that has emerged from this line
of development
\cite{Grishchuk,AV82,HH,L1,Rubakov1,V84,ZelStar} is
that a small closed universe can spontaneously nucleate out of
nothing, where by `nothing' I mean a state with no classical space and
time.  The cosmological wave function can be used to calculate the
probability distribution for the initial configurations of the
nucleating universes.  Once the universe nucleates, it is expected to
go through a period of inflation, 
driven by the energy of a false vacuum.  The vacuum energy
is eventually thermalized, inflation ends, and from then on the
universe follows the standard hot cosmological scenario.  Inflation is
a necessary ingredient in this kind of scheme, since it gives the only
way to get from the tiny nucleated universe to the large universe we
live in today.

The wave function of the universe $\psi$ satisfies the Wheeler-DeWitt
equation, 
\beq 
{\cal H}\psi=0, 
\eeq 
which is analogous to the
Schrodinger equation in ordinary quantum mechanics.  To solve this
equation, one has to specify some boundary conditions for $\psi$.  In
quantum mechanics, the boundary conditions are determined by the
physical setup external to the system.  But since there is nothing
external to the universe, it appears that boundary conditions for the
wave function of the universe should be postulated as an independent
physical law.  The possible form of this law has been debated for
nearly 20 years, and one can hope that it will eventually be derived
from the fundamental theory.

Presently, there are at least three proposals on the table: the
Hartle-Hawking wave function \cite{HH,Hawking86}, the Linde wave
function \cite{L1}, and the tunneling wave function \cite{V84,V86}.

\section{The tunneling wave function}

To introduce the tunneling wave function, let us consider a very
simple model of a closed Friedmann-Robertson-Walker universe filled with a
vacuum of constant energy density $\rho_v$ and some radiation.  The
total energy density of the universe is given by
\beq
\rho=\rho_v+\epsilon/a^4,
\label{1}
\eeq
where $a$ is the scale factor and $\epsilon$ is a constant
characterizing the amount of radiation.  
The evolution equation for $a$ can be
written as
\beq
p^2+a^2-a^4/a_0^2=\epsilon.
\label{2}
\eeq
Here, $p=-a{\dot a}$ is the momentum conjugate to $a$ and 
$a_0=(3/4)\rho_v^{-1/2}$.

\begin{figure}[b!] 
\centerline{
\epsfysize 5 cm \epsffile{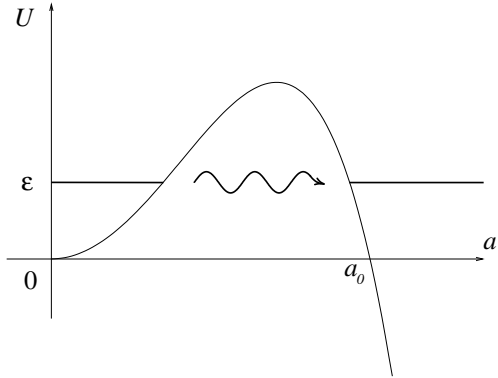}
}
\vspace{10pt}
\caption{The potential for the scale factor in Eq.(\ref{2}).  Instead
of recollapsing, the universe can tunnel through the potential
barrier to the regime of unbounded expansion.}
\label{fig1}
\end{figure}
 
Eq.(\ref{2}) is identical to that for a ``particle'' of energy
$\epsilon$ moving in a potential $U(a)=a^2-a^4/a_0^2$.  For
sufficiently small $\epsilon$, there are two types of classical trajectories.
The universe can start at $a=0$, expand to a maximum radius $a_1$ and
then recollapse.  Alternatively, it can contract from 
infinite size, bounce at a minimum radius $a_2$ and then re-expand
(see Fig. 1).  But in quantum cosmology there is yet another
possibility.  Instead of recollapsing, the universe can
tunnel through the potential barrier to the regime of unbounded
expansion.  The semiclassical tunneling probability
can be estimated as
\beq
{\cal P}\sim\exp\left(-2\int_{a_1}^{a_2} |p(a)|da\right).
\eeq
It is interesting that this probability does not vanish in the limit
of $\epsilon\to 0$, when there is no radiation and 
the size of the initial universe shrinks to
zero.  We then have tunneling from {\it nothing} to a closed universe
of a finite radius $a_0$; the corresponding probability is
\beq
{\cal P}\sim\exp\left(-2\int_0^{a_0} |p(a)|da\right)
=\exp\left(-{3\over{8\rho_v}}\right).
\label{probab}
\eeq
The tunneling approach to quantum cosmology assumes that 
our universe originated in a tunneling event of this kind.  Once it
nucleates, the universe immediately begins a de Sitter inflationary
expansion.

The Wheeler-DeWitt equation for our simple model can be obtained by
replacing the momentum $p$ in (\ref{2}) by a differential
operator\footnote{Here and below I disregard the ambiguity associated
with the ordering of non-commuting factors $a$ and $d/da$.  This
ambiguity is unimportant in the semiclassical domain, which we shall be
mainly concerned with in this paper.}, $p\to -id/d a$,
\beq
\left({d^2\over{da^2}}-a^2+{a^4\over{a_0^2}}\right)\psi(a)=0.
\label{WDW}
\eeq
This equation has outgoing and ingoing wave solutions corresponding to
expanding and contracting universes in the classically allowed range
$a>a_0$ and exponentially growing and decaying solutions in the
classically forbidden range $0<a<a_0$.
The boundary condition that selects the tunneling wave function
requires that $\psi$ should include only an outgouing wave at
$a\to\infty$. The under-barrier wave function is then a linear
combination of the growing and decaying solutions.  The two solutions
have comparable magnitudes near the classical turning point, $a=a_0$,
but the decaying solution dominates in the rest of the under-barrier
region.

The tunneling probability can also be expressed in the language of
instantons.  The nucleated universe after tunneling is described by de
Sitter space, and the under-barrier evolution can be semiclassically
represented by the Euclideanized de Sitter space.  This de Sitter
instanton has the geometry of a four-sphere.  By matching it with the
Lorentzian de Sitter at $a=a_0$ we can symbolically represent
\cite{AV82} the origin of the universe as in Fig.2.  For `normal'
quantum tunneling (without gravity), the tunneling probability ${\cal
P}$ is proportional to $\exp (-S_E)$, where $S_E$ is the instanton
action.  In our case,
\begin{equation}
S_E = \int d^4 x \sqrt{-g}\left(-{R\over{16\pi G}}+\rho_v\right) =
-2\rho_v \Omega_4 = -3/8G^2\rho_v ,
\label{eucaction}
\end{equation}
where $R =32\pi G\rho_v$
is the scalar curvature, and $\Omega_4 = (4\pi^2 /3)a_0^4$ is the volume of 
the four-sphere.  Hence, I concluded in Ref.\cite{AV82} that
${\cal P} \propto \exp (3/8G^2\rho_v)$.

\begin{figure}
\epsfysize 5 cm \epsffile{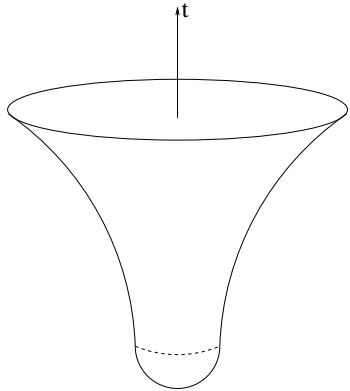}
\caption{A schematic representation of the birth of
inflationary universe.}
\end{figure}

I resist the temptation to call this `the greatest mistake of my
life', but I did change my mind on this issue \cite{V84}.  I now
believe that the correct answer is given by Eq.(\ref{probab}).  In
theories with gravity, the Euclidean action is not positive-definite
and the naive answer no longer applies.  Similar conclusions were
reached independently by Linde \cite{L1}, Rubakov
\cite{Rubakov1} and by Zeldovich and Starobinsky \cite{ZelStar}.  But the
story does not end there.  Not everybody believes that my first answer
was a mistake.  In fact, Stephen and his collaborators believe that I
got it right the first time around and that it is now that I am making
a mistake.  I shall return to this ongoing debate later in this paper.

In the general case, the wave function of the universe is defined on
superspace, which is the space of all 3-dimensional geometries and
matter field configurations,
$\psi [g_{ij}({\bf x}), \varphi ({\bf x})]$,
where $g_{ij}$ is the 3-metric, and matter fields are represented by a
single field $\varphi$.  The tunneling boundary condition can be extended
to full superspace by requiring that $\psi$ should include only
outgoing waves at the boundary of superspace, except the part of the
boundary corresponding to vanishing 3-geometries (see \cite{V86,V94}
for more details).

Alternatively, the tunneling wave function can be defined as a path
integral
\beq
\psi_T(g,\varphi)=\int_\emptyset^{(g,\varphi)}e^{iS},
\label{psiT}
\eeq
where the integration is over paths interpolating between a vanishing
3-geometry $\emptyset$ (`nothing') and $(g,\varphi)$.  I argued
in \cite{V94} that this definition is equivalent to the tunneling
boundary condition in a wide class of models.

Note that
according to the definition (\ref{psiT}), $\psi_T$ is strictly not a wave
function, but rather a propagator
\beq
\psi_T(g,\varphi)=K(g,\varphi |\emptyset),
\eeq
where $K(g,\varphi |g',\varphi')$ is given by the path integral (\ref{psiT})
taken over Lorentzian histories interpolating between $(g',\varphi')$ and $(g,\varphi)$.
One expects, therefore, that $\psi_T$ should generally be singular at
$g=\emptyset$.  In simple minisuperspace models this singularity was
noted in my paper \cite{V86} and also by Kontoleon and Wiltshire
\cite{PW} (who regarded it as an undesirable feature of $\psi_T$).

I should also mention an interesting issue raised by Rubakov
\cite{Rubakov1,Rubakov} who argues that cosmic tunneling of the type 
illustrated in Fig.~1 results in a catastrophic particle production
and in a breakdown of the semiclassical approximation.  This
conclusion is reached using a Euclidean formalism in which particles
are defined by an instantaneous diagonalization of the Hamiltonian.
This is a rather unconventional approach and I am not convinced that
`particles' defined in this way are the same particles that observers
will detect when the universe emerges from under the barrier.  If
Rubakov is right, then the tunneling wave function cannot be obtained
as a limit of tunneling from a small initial universe in a generic
quantum state when the size of that universe goes to zero.  (There is
no dispute that for a particular quantum state, corresponding to the
de Sitter invariant Bunch-Davis vacuum, there is no catastrophic
particle production and the semiclassical approximation is well
justified \cite{VV88,Rubakov}.)  This issue requires further study.

At present, the general definitions of the tunneling wave function
remain largely formal, since we do not know how to solve the
Wheeler-DeWitt equation\footnote{Note that the Wheeler-DeWitt equation
applies assuming that the topology of the universe is fixed.  Possible
modification of this equation accounting for topology change have been
discussed in \cite{V94}.  Here, I have disregarded topology change, assuming
the simplest $S_3$ topology of the universe.  For a discussion of
topology-changing processes in quantum gravity, see Fay Dowker's
contribution to this volume.} or how to calculate the path integral
(\ref{psiT}), except for simple models (and small perturbations about
them), or in the semiclassical limit.  A promising recent
development is the work by Ambjorn, Jurkiewicz and Loll \cite{Ambjorn}
who developed a Lorentzian path integral approach to quantum gravity.
It would be interesting to see an application of this approach to
the tunneling problem.

\section{Alternative proposals for the wave function}

I shall now comment on the other proposals for the wave function of
the universe. 

\subsection{The DeWitt wave function}

I should first mention what I believe was the first
such proposal, made by DeWitt in his 1967 paper \cite{DeWitt}.  DeWitt
suggested that the wave function should vanish for the vanishing scale
factor, 
\beq
\psi_{DW} (a=0)=0.
\label{DW}
\eeq
The motivation for this is that $a=0$ corresponds to the cosmological
singularity, so (\ref{DW}) says that the probability for the
singularity to occur is zero.

The boundary condition (\ref{DW}) is easy to satisfy in a
minisuperspace model with a single degree of freedom, but in more
general models it tends to give an identically vanishing wave
function.  No generalizations of the DeWitt boundary condition
(\ref{DW}) have yet been proposed.

\subsection{The Hartle-Hawking wave function}

The Hartle-Hawking wave function is expressed as a path integral over
compact Euclidean geometries bounded by a given 3-geometry $g$,
\beq
\psi_{HH}(g,\varphi)=\int^{(g,\varphi)}e^{-S_E}.
\label{HH}
\eeq
The Euclidean rotation of the time axis,
$t\to i\tau$,
is often used in quantum field theory because it improves the
convergence of the path integrals.  However, in quantum gravity the
situation is the opposite.  The gravitational part of the Euclidean
action $S_E$ is unbounded from below, and the integral (\ref{HH}) is
badly divergent.  One can attempt to fix
the problem by additional contour rotations, extending the path
integral to complex metrics.  However, the space of complex
metrics is very large, and no obvious choice of integration contour
suggests itself as the preferred one \cite{HalHar}.

In practice, one assumes that the dominant contribution to the path
integral is given by the stationary points of the action and evaluates
$\psi_{HH}$ simply as $\psi_{HH}\sim e^{-S_E}$.  For our simple model,
$S_E= -3/8\rho_v$ and the nucleation probability is 
${\cal P}\sim\exp(+3/8\rho_v)$.  The wave function $\psi_{HH}(a)$ for
this model has only the growing solution under
the barrier and a superposition of ingoing and outgoing waves with
equal amplitudes in the classically allowed region.  This wave
function appears to describe a contracting and re-expanding universe.

It is sometimes argued \cite{Hawking86,Rubakov} that changing
expansion to contraction does not do anything, as long as the
directions of all other physical processes are also reversed.  So if
the ingoing and outgoing waves are CPT conjugates of one another, they
may both correspond to expanding universes, provided that the internal
direction of time is determined as that in which the entropy
increases.  I would like to note that this issue is clarified in
models where the universe is described by a brane propagating in an
infinite higher-dimensional bulk space \cite{Davidson,Bucher,DGP}.
The nucleation of the universe then appears as bubble nucleation
from the point of view of the bulk observer.  In such models, there is
an extrinsic bulk time variable, and the interpretation of incoming
and outgoing waves is unambiguous \cite{Ruben}.  The tunneling wave
function appears to be the only correct choice in this case.

\subsection{The Linde wave function}

Linde suggested that the wave function of the universe is given by a
Euclidean path integral like (\ref{HH}), but with the Euclidean time rotation
performed in the opposite sense,
$t\to +i\tau$,
yielding
\beq
\psi_L=\int^{(g,\varphi)}e^{+S_E}.
\label{psiL}
\eeq
For our simple model, this wave function gives the same nucleation
probability (\ref{prob}) as the tunneling wave function.

The problem with this proposal is that the Euclidean action is also
unbounded from above and, once again, the path integral is divergent.
If one regards Eq.(\ref{psiL}) as a general definition that applies
beyond the simple model (something that Linde himself never
suggested), then the divergence is even more disastrous than in the
Hartle-Hawking case, because now all integrations over matter fields
and over inhomogeneous modes of the metric are divergent.  This
problem of the (extended) Linde's wave function makes it an easy
target, and I suspect it is for this reason that Stephen likes to
confuse $\psi_L$ and $\psi_T$ and refers to both of them as ``the
tunneling wave function''.  In fact, the two wave functions are quite
different, even in the simplest model \cite{debate}.  The Linde wave function
includes only the decaying solution under the barrier and a
superposition of ingoing and outgoing modes with equal amplitudes
outside the barrier.

Using Stephen's expression, I think it would be fair to say that at
present none of us ``has a wave function''.  All four proposals are
well defined only in simple minisuperspace models or in the
semiclassical approximation.  So they are to be regarded only as
prototypes for future work in this area, and not as well defined
mathematical objects.

\section{Semiclassical probabilities}

Quantum cosmology is based on quantum gravity and shares all of its
problems.  In addition, it has some extra problems which arise when
one tries to quantize a closed universe.  The first problem stems from
the fact that $\psi$ is independent of time.  This can be understood
\cite{DeWitt} in the sense that the wave function of the universe
should describe everything, including the clocks which show time.  In
other words, time should be defined intrinsically in terms of the
geometric or matter variables.  However, no general prescription has
yet been found that would give a function $t(g_{ij},\varphi)$ that
would be, in some sense, monotonic.  

A related problem is the definition of probability.  Given a wave
function $\psi$, how can we calculate probabilities?  There was some
debate about this in the 1980's, but now it seems that the only
reasonable definition that we have is in terms of the conserved
current of the Wheeler-DeWitt equation \cite{DeWitt,Misner,V89}.
The Wheeler-DeWitt equation can be symbolically written in the form
\begin{equation}
(\nabla^2 -U)\psi =0,
\label{Kleingordon}
\end{equation}
which is similar to the Klein-Gordon equation.  Here, $\nabla^2$ is
the superspace Laplacian and the `potential' $U$ is a functional of
$g_{ij}$ and $\varphi$.  (We shall not need explicit forms of
$\nabla^2$ and $U$.)  This equation has a conserved current
\begin{equation}
J=i(\psi^*\nabla\psi -\psi\nabla\psi^*),~~~~~~~\nabla\cdot J=0.
\label{current}
\end{equation}
The conservation is a useful property, since we want probability to be
conserved.  But one runs into the same problem as with the Klein-Gordon
equation: the probability defined using (\ref{current}) is not
positive-definite.  Although we do not know how to solve these
problems in general, they can both be solved in the semiclassical
domain.  In fact, it is possible that this is all we need.
 
Let us consider the situation when some of the variables $\{ c\}$ describing
the universe behave classically, while the rest of the variables $\{ q \}$ must
be treated quantum-mechanically.  Then the wave function of the universe can be
written as a superposition
\begin{equation}
\psi =\sum_k A_k(c)e^{iS_k(c)}\chi_k (c,q) \equiv\sum_k\psi_k^{(c)}\chi_k,
\label{WKB}
\end{equation}
where the classical variables are described by the WKB wave functions
$\psi_k^{(c)}
=A_ke^{iS_k}$.  In the semiclassical regime, $\nabla S$ is large, and 
substitution
of (\ref{WKB}) into the Wheeler-DeWitt equation (\ref{Kleingordon}) yields the
Hamilton-Jacobi equation for $S(c)$,
\begin{equation}
\nabla S\cdot\nabla S +U=0.
\label{hamjac}
\end{equation}
The summation in (\ref{WKB}) is over different solutions of this
equation.  
Each solution of (\ref{hamjac}) is a classical action describing a congruence
of classical trajectories (which are essentially the gradient curves of $S$). 
Hence, a semiclassical wave function $\psi_c =Ae^{iS}$ describes an ensemble of
classical universes evolving along the trajectories of $S(c)$.  A probability
distribution for these trajectories can be obtained using the conserved 
current 
(\ref{current}).  Since the variables $c$ behave classically, these 
probabilities
do not change in the course of evolution and can be thought of as probabilities
for various initial conditions.  The time variable $t$ can be defined as any
monotonic parameter along the trajectories, and it can be shown 
\cite{DeWitt,V89} that in this case 
the corresponding component of the current $J$ is
non-negative, $J_t \geq 0$.  Moreover, one finds \cite{LapRub,HalHaw,Banks85}
that the `quantum' wave function $\chi$ satisfies the usual Schrodinger
equation,
\begin{equation}
i\partial\chi /\partial t ={\cal H}_\chi \chi
\end{equation}
with an appropriate Hamiltonian ${\cal H}_\chi$.  
Hence, all the familiar physics is
recovered in the semiclassical regime.

This semiclassical interpretation of the wave function $\psi$ is valid to the
extent that the WKB approximation for $\psi_c$ is justified and the
interference between different terms in (\ref{WKB}) can be neglected. 
Otherwise, time and probability cannot be defined, suggesting that the wave
function has no meaningful interpretation.  In a universe where no object
behaves classically (that is, predictably), no clocks can be constructed, no
measurements can be made, and there is nothing to interpret.  It would
be interesting, however, to investigate the effects of small
corrections to the WKB form of the wave functions and of non-vanishing
interference.

\section{Comparing different wave functions}

To see what kind of cosmological predictions can be obtained from
different wave functions, one needs to consider an extension of the
minisuperspace model (\ref{WDW}).  Instead of a constant vacuum energy
$\rho_v$, one introduces an inflaton field $\varphi$ with a potential
$V(\varphi)$.  The Wheeler-DeWitt equation for this two-dimensional
model can be approximately solved assuming that $V(\varphi)$ is a
slowly-varying function and is well below the Planck density.

After an appropriate rescaling of the scale factor $a$ and the scalar field
$\varphi$, the Wheeler-DeWitt equation can be written as
\begin{equation}
\left[ {\partial^2 \over{\partial a^2}}-{1\over{a^2}}{\partial^2
\over{\partial\varphi^2}}-U(a,\varphi)\right]\psi(a,\varphi)=0,
\label{WDWaphi}
\end{equation}
where
\begin{equation}
U(a,\varphi)=a^2[1-a^2V(\varphi)].
\label{potentialaphi}
\end{equation}
With the above assumptions, one finds \cite{V88} that
Hartle-Hawking and tunneling solutions of this equation are given
essentially by the same expressions as for the simple model (\ref{WDW}), but
with $\rho_v$ replaced by $V(\varphi)$.  The only difference is that the wave
function is multiplied by a factor $C(\varphi)$, such that 
$\psi(a,\varphi)$ becomes
$\varphi$-independent in the limit $a\to 0$ (with $|\varphi |<\infty$).

The initial state of the nucleating universe in this model is
characterized by the value of the scalar field $\varphi$, with the
initial value of $a$ given by $a_0(\varphi)=V^{-1/2}(\varphi)$.  The
probability distribution for $\varphi$ can be found using the
conserved current (\ref{current}).  For the tunneling wave function
one finds
\begin{equation}
{\cal P}_T(\varphi)\propto\exp \left( -{3\over{8G^2V(\varphi)}}\right),
\label{tunnelprob}
\end{equation}
This is the same as
Eq.(\ref{probab}) with $\rho_v$ replaced by $V(\varphi)$.  

Eq.(\ref{tunnelprob}) can be interpreted as the probability
distribution for the initial values of $\varphi$ in the ensemble of
nucleated universes.  The highest probability is obtained for the
largest values of $V(\varphi)$ (and smallest initial size).  Thus, the
tunneling wave function `predicts' that the universe is most likely to
nucleate with the largest possible vacuum energy.  This is just the
right initial condition for inflation.  The high vacuum energy drives
the inflationary expansion, while the field $\varphi$ gradually `rolls
down' the potential hill, and ends up at the minimum with
$V(\varphi)\approx 0$, where we are now.

The Hartle-Hawking wave function gives a similar
distribution, but with a crucial difference in sign,
\begin{equation}
{\cal P}_H(\varphi)\propto\exp \left(+{3\over{8G^2V(\varphi)}}\right).
\label{HHprob}
\end{equation}
This is peaked at the smallest values of
$V(\varphi)$.  Thus the Hartle-Hawking wave function tends to predict
initial conditions that disfavor inflation.  There has been much
discussion of this point in the literature 
\cite{V88,HawPage,Grish,KamBarv,Page}, 
but as I will explain in Section 8, the eternal nature of inflation
makes this distinction between the wave functions rather irrelevant.

\section{Do we need quantum cosmology?}

The status of quantum cosmology is closely related to that of eternal
inflation, and I am going to discuss this relation in the rest of the
paper.

A very generic feature of inflation is its future-eternal character
\cite{V83,L2,Starob}.  The evolution of the inflaton field $\varphi$ is
influenced by quantum fluctuations, and as a result thermalization
does not occur simultaneously in different parts of the universe.  One
finds that, at any time, the universe consists of post-inflationary,
thermalized regions embedded in an inflating background.
Thermalized regions grow by annexing adjacent
inflating regions, and new thermalized regions are constantly formed
in the midst of the inflating sea.  At the same time, the inflating
regions expand and their combined volume grows exponentially with time.
It can be shown that the inflating regions forms a
self-similar fractal of dimension somewhat smaller than 3
\cite{Aryal,Winitzki}. 

Given this picture, it is natural to ask if the universe could also be
past-eternal.  If it could, we would have a model of an infinite, 
eternally inflating universe without a beginning.  We would then need 
no initial or boundary conditions for the universe, and quantum
cosmology would arguably be unnecesary. 

This possibility was discussed in the early days of inflation, but it
was soon realized \cite{Steinhardt83,Linde83} that the idea could not
be implemented in the simplest model in which the inflating universe
is described by an exact de Sitter space.  The reason is that in the
full de Sitter space, exponential expansion is preceeded by an
exponential contraction.  If thermalized regions were allowed to form
all the way to the past infinity, they would rapidly fill the space,
and the whole universe would be thermalized before the inflationary
expansion could begin.  

More recently, general theorems were proved \cite{Borde1}, using the
global techniques of Penrose and Hawking, where it was shown that
inflationary spacetimes are geodesically incomplete to the past.
However, it is now believed \cite{Borde2,GVW} that one of the key
assumptions made in these theorems, the weak energy condition, is
likely to be violated by quantum fluctuations in the inflating parts
of the universe.  This appears to open the door again to the
possibility that inflation, by itself, can eliminate the need for
initial conditions.  Now I would like to report on a new theorem,
proved in collaboration with Arvind Borde and Alan Guth \cite{BGV},
which appears to close that door completely.  (For a more detailed
discussion of the theorem and an outline of the proof, see Alan Guth's
contribution to this volume.)

The theorem assumes that (i) the spacetime is globally expanding and
(ii) that the expansion rate is bounded below by a positive constant,
\beq
H>H_{min}\geq 0.
\label{Hbound}
\eeq
The theorem states that any spacetime with these properties is past
geodesically incomplete.  
Both of the above conditions need to be spelled out.  
  
It is important to realize that expansion and contraction are not
local properties of spacetime.  Rather, they refer to the relative
motion of comoving observers filling the spacetime, with observers
being described by a congruence of timelike geodesics.  The global
expansion condition requires that the spacetime can be filled by an
expanding congruence of geodesics.\footnote{This formulation of the
global expansion condition may be somewhat too restrictive.
Congruences of geodesics tend to develop caustics and often cannot be
globally defined.  A weaker form of the condition, which is still
sufficient for the proof of the theorem, requires that an expanding
congruence satisfying (\ref{Hbound}) can be continuously defined along
a past-directed timelike or null geodesic.  Members of the
congruence may cross or focus away from that geodesic.}  This
condition is meant to exclude spacetimes like de Sitter space (which
can be said to have a globally contracting phase).

The Hubble expansion rate $H$ is defined as usual, as the relative
velocity divided by the distance, with all quantities measured in the
local comoving frame.  Since we do not assume any symmetries of
spacetime, the expansion rates are generally different at different
places and in different directions.  The bound (\ref{Hbound}) is
assumed to be satisfied at all points and in all directions.  This is
a very reasonable requirement in the inflating region of spacetime.

The theorem is straightforwardly extended to higher-dimensional
models.  For example, in Bucher's model \cite{Bucher}, brane
worlds are created in collisions of bubbles nucleating in an inflating
higher-dimensional bulk spacetime.  Our theorem implies that the
inflating bulk cannot be past-eternal.  Another example is the cyclic
brane world model of Steinhardt and Turok \cite{ST}.  One of the two
branes in this model is globally expanding and thus should have a past
boundary.

This is good news for quantum cosmology.  It follows from the theorem
that the inflating region has a boundary in the past, and some new
physics (other than inflation) is necessary to determine the
conditions at that boundary.  Quantum cosmology is the prime candidate
for this role.  The picture suggested by quantum cosmology is that the
universe starts as a small, closed 3-geometry and immediately enters
the regime of eternal inflation, with new thermalized regions being
constantly formed.  In this picture, the universe has a beginning, but
it has no end.

\section{Is quantum cosmology testable?}

There are also some bad news.  In the course of eternal inflation, the
universe quickly forgets its initial conditions.  Since the number of
thermalized regions to be formed in an eternally inflating universe is
unbounded, a typical observer is removed arbitrarily far from the
beginning, and the memory of the initial state is completely erased.
This implies that any predictions that quantum cosmology could make
about the initial state of the universe cannot be tested
observationally.  All three proposals for the wave function of the
universe are therefore in equally good agreement with observations, as
well as a wide class of other wave functions -- as long as they give a
non-vanishing probability for eternal inflation to start
\cite{LLM}.\footnote{There is, arguably, a much wider class of wave functions
which describe highly excited states of the fields, but these will
generally exhibit no quasiclassical behavior and will not, therefore,
allow for the existence of observers \cite{Hartle}.}  

The only case that requires special
consideration is when there are some constants of nature, $\alpha_j$,
which are constant within individual universes, but can take different
values in different universes of the ensemble.  (One example is the
cosmological constant in models where it is determined by a four-form
field.)  In this case, the memory of the initial state is never erased
completely, since the values of $\alpha_j$ are always equal to their
initial values.  One might hope that probabilistic predictions for the
values of $\alpha_j$ could be derived from quantum cosmology and
could, in principle, be tested observationally.  Unfortunately, this
prospect does not look very promising either.

Quantum cosmology can give us the probability distribution
$P_{nucl}(\alpha)$ for a universe to nucleate with given values of
$\alpha_j$.  In other words, this is the probability for a universe
arbitrarily picked in the ensemble to have this set of values.
To get the probability of observing these values, it should be
multiplied by the average number of independent observers,
$N(\alpha)$, that will evolve in such a universe \cite{V95},
\beq
P_{obs}(\alpha)\propto P_{nucl}(\alpha)N(\alpha).
\label{prob}
\eeq

The number of observers in each eternally inflating universe grows
exponentially with time,
\beq
N(\alpha;t) =B(\alpha) \exp[\chi(\alpha)t].
\eeq
The prefactor $B(\alpha)$ depends on the details of the biochemical
processes, and at present we have no idea how to calculate it.  But the
rate of growth $\chi$ is determined by the growth of thermalized
volume and can be found as an eigenvalue of the
Fokker-Planck operator, as discussed in Refs. \cite{Starob,LLM}.  It is
independent of biology, but
generally depends on $\alpha_j$.  This suggests that the most probable
values of $\alpha_j$ should be the ones maximizing the expansion rate
$\chi(\alpha)$.  As time goes on, the number of observers  in universes
with this preferred set of $\alpha_j$ gets larger than the competition
by an arbitrarily large factor.  In the limit $t\to\infty$, this set
has a 100\% probability, while the probability of any other values is
zero,
\beq
P_{obs}(\alpha)\propto \delta(\alpha-\alpha_*),
\label{p}
\eeq
where $\chi(\alpha_*)={\rm max}$.  We thus see that the probability of
observing the constants $\alpha_j$ is determined entirely by the
physics of eternal inflation and is independent of the nucleation
probability $P_{nucl}(\alpha)$ -- as long as $P_{nucl}(\alpha_*)\not= 0$.

The situation I have just described is somewhat clouded by the problem
of gauge-dependence.  The problem is that the expansion rate
$\chi(\alpha)$ and the values of $\alpha_j$ maximizing this rate
depend on one's choice of the time coordinate $t$ \cite{LLM}.  Time in General
Relativity is an arbitrary label, and this gauge dependence casts
doubt on the meaningfulness of the probability (\ref{p}).  We now have
some proposals on how this problem can be resolved in a single
eternally inflating universe \cite{V98,GarVil}, but comparing the
numbers of observers in disconnected eternally inflating universes
still remains a challenge.

I can think of two possible responses to this situation.  (i) There may
be some preferred, on physical grounds, choice of the time variable
$t$, which should be used in this case for the calculation of
probabilities.  For example, one could choose the proper time along
the worldlines of comoving observers.  (ii) One can take the point of
view that no meaningful definition of probabilities is possible for
observations in disconnected, eternally inflating universes.  While
this issue requires further investigation, the important point for us
here is that, in either case, an observational test of quantum
cosmology does not seem possible.

Thus, the conclusion is that, sadly, quantum cosmology is not likely
to become an observational science.  However, without quantum
cosmology our picture of the universe is incomplete.  It raises very
intriguing questions of principle and will, no doubt, inspire future
research.  I wish that Stephen continues to lead and challenge us in
this adventure.


\end{document}